# Chirality in magnetic multilayers probed by the symmetry and the amplitude of dichroism in X-ray resonant magnetic scattering

*Jean-Yves Chauleau[1,2], William Legrand[2], Nicolas Reyren[2], Davide Maccariello[2], Sophie Collin[2], Horia Popescu[1], Karim Bouzehouane[2], Vincent Cros[2], Nicolas Jaouen[1], and Albert Fert[2]*

[1] *Synchrotron SOLEIL, L'Orme des Merisiers, 91192, Gif-sur-Yvette, France*
[2] *Unité Mixte de Physique, CNRS, Thales, Univ. Paris-Sud, Université Paris-Saclay, 91767, Palaiseau, France.*

*Chirality in condensed matter is now a topic of the utmost importance because of its significant role in the understanding and mastering of a large variety of new fundamental physicals mechanisms. Versatile experimental approaches, capable to reveal easily the exact winding of order parameters are therefore essential. Here we report X-ray resonant magnetic scattering (XRMS) as a straightforward tool to identify directly the properties of chiral magnetic systems. We show that it can straight-forwardly and unambiguously determine the main characteristics of chiral magnetic distributions: i.e. its chiral nature, the quantitative winding sense (clockwise or counter-clockwise) and its type (Néel/cycloidal or Bloch/helical). This method is model-independent, does not require a-priori knowledge of magnetic parameters and can be applied to any system with magnetic domains ranging from few nanometers (wavelength limited) to several microns. By using prototypical multilayers with tailored magnetic chiralities based on the Co|Pt interface we illustrate the strength of this method.*

Chirality is central to understand many fundamental mechanisms in various domains of physics and chemistry. In condensed matter, a large variety of physical phenomena hinge on the emergence of these complex chiral windings of order parameters, their observation and subsequently their control especially in magnetism and spin-transport at the nanoscale. The ability to probe the nature of these chiral magnetic textures has now become a crucial element of modern magnetism and is therefore essential to gain a deeper understanding of these mechanisms. Spin-polarized scanning tunneling microscopy (SP-STM) revealed that magnetic textures with a cycloidal configuration of the magnetization and Néel domain walls are stabilized in ultra-thin magnetic films (one or a few atomic layers) on heavy metal substrates [1] . It was realized that these magnetic textures are stabilized by Dzyaloshinskii-Moriya (DM) interaction [2,3], ($\sum_{\langle ij \rangle} \boldsymbol{D}_{ij} \cdot \boldsymbol{S}_i \times \boldsymbol{S}_j$, with $\boldsymbol{S}_i$ and $\boldsymbol{S}_j$ two neighboring spins) which is the anti-symmetric analog of the Heisenberg interaction favoring a curling magnetization textures around the DM vector $\boldsymbol{D}$. The DM interaction requires spin-orbit coupling (SOC) and broken inversion symmetry, found either in specific crystalline structures, such as B20 materials [4], or at film interfaces [5,6].

In this letter, we demonstrate that XRMS experiments (Figure 1a) allow us to reveal the actual magnetic textures existing in ultrathin magnetic multilayers with perpendicular anisotropy and large interfacial chiral interaction. Importantly, this determination is straightforward and does not require any assumptions or careful comparison to parameter-dependent micromagnetic simulations. Moreover, the circular dichroism in XMRS enables to identify directly not only the direction but also the sense of the magnetic winding and therefore of the actual sign of DM interaction as we proved by a thorough analysis. This information has a far-reaching impact in particular for spin-orbitronics and spin-orbit torque (SOT) studies. Indeed, SOT is a recent very promising approach to move efficiently domain walls [7] and magnetic skyrmions along magnetic race-tracks and shows therefore a great potential for future spintronic devices. Yet, the detailed texture of the domain walls plays a major role in their motion [8]: the DW high speed is related to its nature (Néel or Bloch) while its direction of motion depends on its chirality [7].  Besides, DM interaction can play an important role for existing technological applications in spintronics already using ultrathin magnetic multilayers with perpendicular magnetic anisotropy such as MRAM or detectors, in which the presence and the impact of non-negligible DM interaction has been largely overlooked (see e.g. [9,10]), emphasizing thus the need to probe its consequences on magnetic textures. XRMS presents several advantages that might soon turn it to a standard way to characterize and quantify the DM interaction. First, XMRS is a non-perturbative technique contrary to the approach based on magnetic domain analysis though magnetic force microscopy (MFM) imaging or domain wall motion. Second, such study is applicable not only to metallic chiral magnets but can be extended to the substantial number of insulating magnetic systems. Third, this experiment is based on the scattering on magnetic domains so sample patterning is not required. Then, this technique is sensitive to thicknesses of only a few nanometers of total magnetic materials (either in a single film or in multilayers) that conveniently can be buried below a few nanometers of materials (e.g. capping). Finally, XMRS experiments can be extended to time resolved experiments using synchrotron [11] or free electron laser [12], which should open new horizons to the exploration of the dynamics of magnetic chirality.

Here, we perform XRMS experiments using circularly polarized x-rays on prototypical Co/Pt based multilayers which have proven to be systems of choice for the emergence of homochiral magnetic distributions and are consequently also prototypical for SOT applications [13,14]. Our magnetic multilayers have been prepared using magnetron sputtering on thermally oxidized silicon substrates (see [15] for details]. Two types of samples composed of five-fold repetition of a trilayer are considered: //[Pt(1nm)/Co(0.8nm)/Ir(1nm)]x5 and its reversed-stacking counterpart, //[Ir(1nm)/Co($t$)/Pt(1nm)]x5. The symbol "//" stands for the substrate and the 10-nm-thick Pt buffer. To avoid oxidation, a 3-nm-thick Pt capping is deposited on top of the multilayers. In these samples, we measure a saturation magnetization $M_S \approx 1$ MA/m and a large effective out-of-plane anisotropy $K_{eff} \approx 0.1$ MJ/m$^3$. According to our previous studies on these systems [16], the mean DM magnitude (arising mainly from the Pt/Co interfaces but also from Co/Ir interfaces) is about 2 mJ/m² . Such DM magnitude is large enough to impose pure Néel DW texture [8], with counterclockwise (CCW) Néel DW structure in case of a thin Co layer deposited on top of Pt, and conversely a clockwise (CW) Néel DW in case of Pt is deposited on Co. Before running the XRMS experiments, magnetic domain configurations have been imaged at room temperature using (MFM) after a demagnetization process using an out-of-plane field. In Figure 1b, we display a characteristic randomly disordered magnetic stripe patterns with a mean period of about 180 ± 30 nm (see the FFT of the MFM image in the inset of Figure 1b).

XRMS experiments have been performed on the two types of multilayer stackings at the SEXTANTS beamline [17] of the SOLEIL synchrotron. They have been performed in reflectivity conditions for circularly left (CL) and right (CR) incident X-ray beam polarizations at the Co $L_3$ edge (photon energy = 778.2 eV) using the RESOXS diffractometer [18]. The diffracted X-rays are collected on a Peltier-cooled square CCD detector covering 6.1° at the working distance of this study. Typical diffracted patterns of the domain structure are displayed in Figures 1c-e. All the images have been geometrically corrected along the $Q_x$-direction in order to account for the projection related to the photon incidence angle θ (of 18.5° in this case). As shown in Fig. 1e, the sum of the images obtained with CR (Fig. 1c) and CL (Fig. 1d) polarized light gives rise to a clear ring-shaped diffraction pattern around the specular beam (blocked by a beamstop to avoid saturation of the CCD) in the reciprocal plane ($Q_x$,$Q_y$). The ring radius, labelled as $q_p$ indicates the in-plane isotropy of the domain sizes with a typical periodicity in real space $p = 2\pi/q_p$. The domain size $p/2$ of 90 ± 9 nm agrees with the one determined by MFM (see Fig. 1b) as the Fourier transform of the MFM image can be directly compared to the XRMS sum image. The circular dichroism (CD) of the scattering signal is defined as ($I_{CL}$-$I_{CR}$)/( $I_{CL}$+$I_{CR}$), where $I_{CL}$ and $I_{CR}$ are the intensities collected by the camera for CL and CR polarizations respectively. The colored map of the diffracted CD in Fig. 1f displays two lobes, one blue and one red, indicating an opposite dichroism sign. The amplitude of the dichroism is rather large, typically 10% of the sum signal, and for sure sufficient to univocally reveal the nature of the DW texture as explained hereafter.

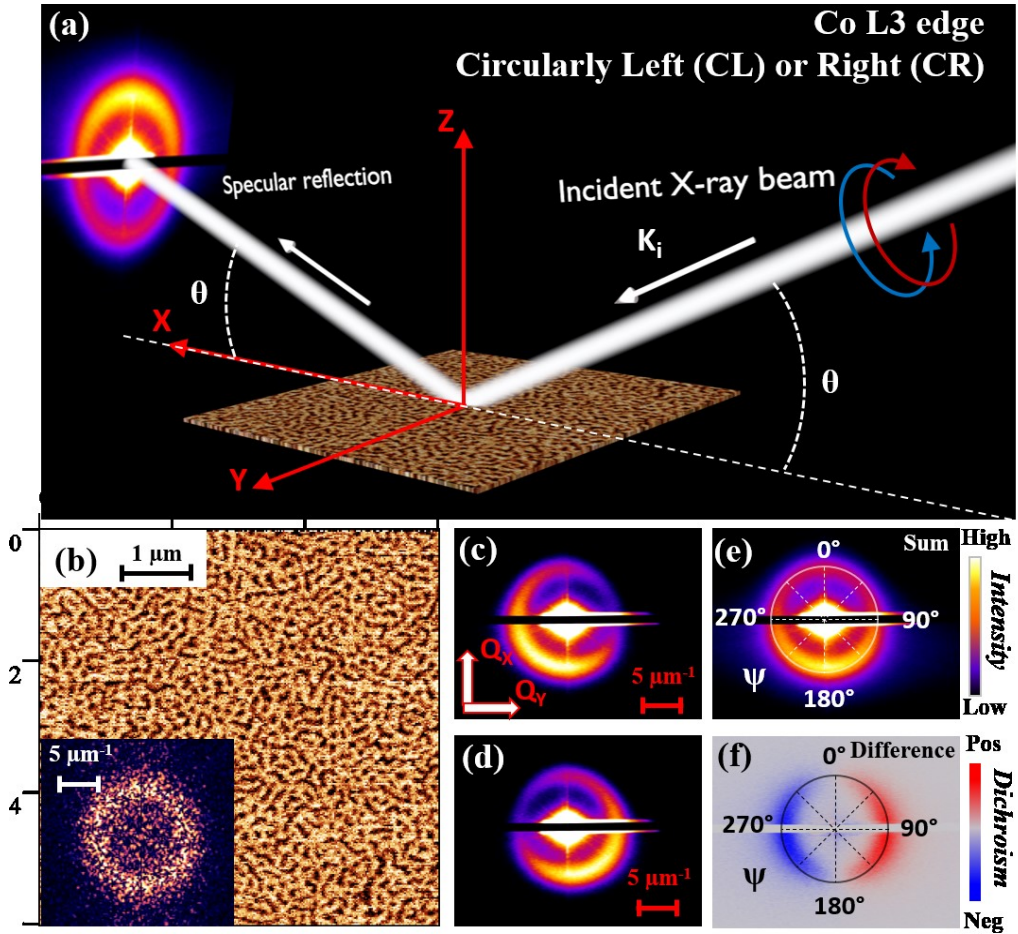

*Figure 1 : a) XRMS measurement principle and experimental configuration. b) 6x6 µm² MFM phase image showing the out-of-plane magnetic contrast of [Ir(1)/Co(0.6)/Pt(1)]x5 multilayer with its corresponding FFT pattern (inset) evidencing a 180 ± 30nm period disordered magnetic stripe pattern . Corresponding diffracted patterns for circularly left, CL (c) and right, CR (d) polarized incident X-ray beam. e) Resulting sum image of the diffracted pattern (CL+CR), confirming that the diffraction corresponds to the magnetic domains observed by MFM. f) Normalized difference image (CL-CR/CL+CR) evidencing a pronounced circular dichroism. Note that the straight black area crossing the diffraction ring is the shadow of a beam-stop used to block the specular reflection.*

Indeed, XRMS has been recently used to characterize magnetic bulk compound hosting skyrmions. In this case, it unambiguously determined the winding number of skyrmions lattices in $Cu_2OSeO_3$ [19,20] through the analysis of the symmetry of the scattering signal. In our work, by leveraging the dichroism in XRMS scattering signal observed with circularly polarized light, we demonstrate how it provides a simple and straightforward novel approach to identify the nature of DW texture, i.e. its character (Néel or Bloch) and the magnetic chirality. We believe that these results, that take inspiration in the study by Dürr *et al.* [21] of the chiral behavior of closure magnetic domains at the surface of 40 nm thick FePd layer, is of high importance given the huge recent interest for this type of magnetic multilayers with large interfacial DM interaction leading to promising applications for a new generation of spin-orbitronic devices [22].

In order to precisely analyze the dichroism in XRMS, we use the following expression of the diffracted intensity $I(\mathbf{Q})$ for a scattering vector $\mathbf{Q} = \mathbf{k}_f - \mathbf{k}_i$ in the kinematical approximation:

$$I(\mathbf{Q}) \propto \left| \sum_n f_n \cdot \exp(i\mathbf{Q} \cdot \mathbf{r}_n) \right|^2$$

where $f_n$ is the resonant scattering amplitude of a single ion at the $r_n$ position in the electron-dipole approximation. In case this ion carries a magnetic moment $\mathbf{m}_n$, different contributions to this scattering amplitude can be distinguished at resonance [23–25] $f_n = f_0 + f_m^1 + f_m^2$ where $f_0$ is the anomalous charge scattering amplitude and $f_m{}^1$ and $f_m{}^2$ are the magnetic resonant scattering amplitudes:

$$f_m^1 \propto -i\,(\hat{\epsilon} \times \hat{\epsilon}') \cdot \boldsymbol{m}_n$$
$$f_m^2 \propto (\hat{\epsilon}' \cdot \boldsymbol{m}_n)(\hat{\epsilon} \cdot \boldsymbol{m}_n)$$

with $\hat{\epsilon}$ and $\hat{\epsilon}'$ the polarization state of the incident and diffracted X-ray beam respectively. Note that while $f_m^1$ scales with **m**, $f_m^2$ scales with **m**$^2$. Considering the extracted characteristic domain period, the magnetic diffracted intensity is mainly related to $f_m^1$. The diffracted intensity for a given incident beam polarization is expressed as follows [26]:

$$I(\boldsymbol{Q}) = \text{Tr}[\tilde{f}_n\, \rho\, \tilde{f}_n^\dagger]$$

where $\tilde{f}_n$ and $\tilde{f}_n^\dagger$ are the Fourier transform of the scattering amplitude $f_n$ and its complex conjugate and $\rho$ is the density matrix of the incident X-ray beam. In the Stoke-Poincaré representation [25] this density matrix for a circularly left or right incident beam is expressed as follows:

$$\rho_{\text{CL}} = \begin{pmatrix} 1 & -i \\ +i & 1 \end{pmatrix} \quad \rho_{\text{CR}} = \begin{pmatrix} 1 & +i \\ -i & 1 \end{pmatrix}$$

In the following, we consider two types of possible magnetic windings, i.e. helicoidal (or Bloch-like) and cycloidal (or Néel-like) depicted in Fig. 2a and 2b respectively, that are the ones expected in the multilayer systems considered in this study. From both these winding configurations, the circular dichroism of their diffraction patterns can be calculated using $(I_{\text{CL}} - I_{\text{CR}})/(I_{\text{CL}} + I_{\text{CR}})$. In Fig. 2c, we present the calculated orthoradial profiles of the normalized circular dichroism for the two possible windings and for each of them for the two possible chiralities, i.e. clockwise (CW) or counterclockwise (CCW). For the case of a helicoidal winding (Bloch-like), the circular dichroism is respectively maximum (for CW) and minimum (for CCW) in positions corresponding to incident beam plane of $\psi = 0$ or 180° and vanishes for $\psi = 90°$ or 270°. On the contrary, for a cycloidal winding (Néel-like), the maximum (for CW) and minimum (for CCW) are obtained for $\psi = 90°$ and 270°, resulting in the rotation of the dichroic diffraction pattern by 90° around the specular beam. Thus, it provides a simple way to determine unambiguously the actual texture of the domain wall from the orientation of the experimentally observed dichroism. In Fig. 2d, we display the orthoradial experimental profile of the circular dichroism we obtained for the //[Ir(1)/Co(0.8)/Pt(1)]x5 multilayer measured at an incidence angle $\theta = 17.5°$. We emphasize that this angle corresponds to the first multilayer peak where the dichroic signal is enhanced. From the comparison with the predicted profile (yellow curve in Fig. 2d), and notably the positions of the maxima and minima, we can directly assert that the magnetic winding in our Ir/Co/Pt multilayers is of Néel type. Hence, it confirms by a direct measurement and without assumptions, that the DW texture is Néel (CW or CCW) as already known for this multilayer system with additive interfacial DM interaction either indirectly through analysis of the skyrmion size and comparison with micromagnetic simulations [15] or directly through Lorentz transmission electron microscopy [27]. Other methods have already been utilized to access to the texture of the DW or skyrmions such as spin-polarized STM [1,28], scanning nitrogen-vacancies-magnetometry (NV) [29,30] or spin-polarized low-energy electron microscopy (SPLEEM) [31]. On the other hand, DM interaction has also been probed by spin wave spectroscopy techniques such as Brillouin light scattering (BLS) [32] or time-resolved Kerr microscopy [33]. Once the type of winding is determined, the chirality (CW or CCW) of the Néel texture is usually indirectly deduced, for example, from DW dynamics and micromagnetic comparison [36], is straight-forwardly assessed by XRMS. Indeed, from Fig. 2d, we conclude that the magnetic texture for sample //[Ir(1)/Co(0.8)/Pt(1)]x5 corresponds to a fixed CW Néel DW chirality. Finally, the global circular dichroism pattern (changing sign only twice) allows to corroborate that the overall topological winding number of the entire magnetic distribution is *N*=1 [19].

We notice that beyond this qualitative comparison, we observe a deviation between the measurement and the calculated curves in Fig. 2d outside the angles corresponding to maxima and minima. We attribute it to the use of the kinematical approximation in our simulation that is known not to be strictly valid in the soft x-ray range [35] but this does not affect the main conclusion.

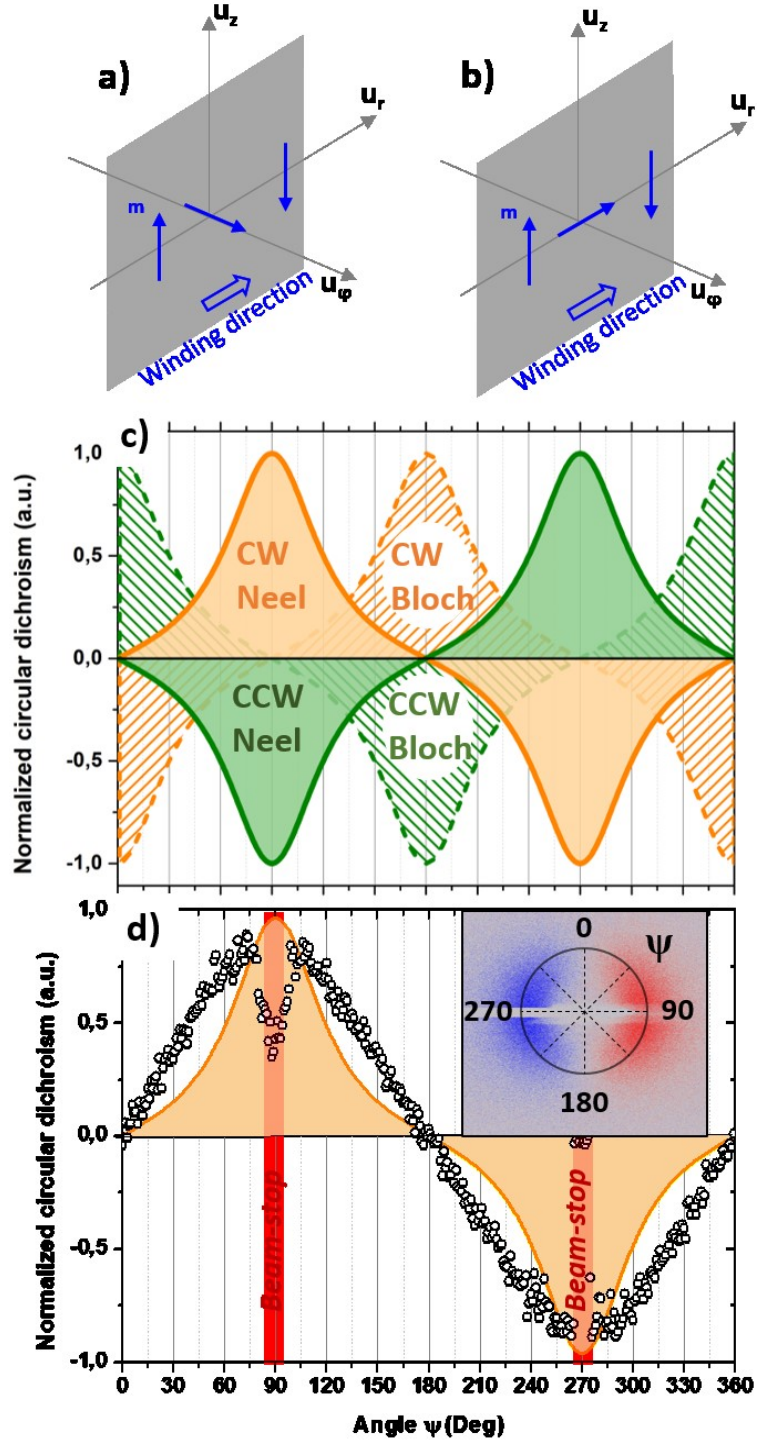

*Figure 2 : (a) diagram of a Bloch (helicoidal) winding and of a (b) Néel (cycloidal) winding. (c) Calculated orthoradial profile of the circular dirchroism of different magnetic textures: clockwise (CW, orange) and counter-clockwise (CCW, green) Bloch (dashed) and Néel (solid) windings for an incident angle ϑ=17.5°. (d) Comparison with the experimental circular dichroism orthoradial profile of a [Ir(1)/Co(0.8)/Pt(1)]x5 multilayer for the same incident angle ϑ=17.5° (dots). Note that the red rectangles indicate the position of the beamstop.*

In order to experimentally confirm the origin of this dichroic contrast in the diffraction maps, we compare directly in Fig. 3, two samples having the same multilayer constitution but opposite stacking: //[Ir(1)/Co(0.8)/Pt(1)]x5 (top of Fig. 3) or //[Pt(1)/Co(0.8)/Ir(1)]x5 (bottom of Fig. 3). In these two multilayers, we expect a reversed direction of DM vector as predicted from the shape of the interfacial origin of the DM interaction [5,16] in bilayers or multilayers with magnetic material in contact with heavy materials. In Fig. 3b and 3e, we display the MFM images obtained after demagnetization under perpendicular field on each sample prior to the XRMS experiments. For both samples, we find a maze domain configuration, with a difference in the mean domain width (see FFT in the insets of Fig. 3b and Fig. 3e) due to some (quasi-unavoidable) non-negligible difference in the magnetic parameters of the two multilayers, such as the perpendicular magnetic anisotropy, saturation magnetization and DM interaction. However, despite these differences, it can be directly and unambiguously concluded from the dichroism XRMS diagram (Fig. 3c and 3f) that both the //Ir/Co/Pt and //Pt/Co/Ir multilayers have the same pure Néel domain wall configuration but with opposite winding signs. Indeed, they both present a complete disappearance of the dichroism for $\psi = 0$ or $180°$ but with a clear reversal of the sign of the dichroism. Thus, a simple XRMS map taking a few minutes (the dichroism is so large that even a single circular polarity might be enough to conclude) directly reveals that a system with Co on top of Pt has a CCW Néel DW configuration, while for Co below Pt, a CW Néel DW is found, as expected from theoretical predictions for bilayers. The result of energy minimization using micromagnetic simulations using MuMax3 code [36], using our film parameters, confirms opposite Néel DW spin textures in the two stackings (Fig. 3a and 3d).

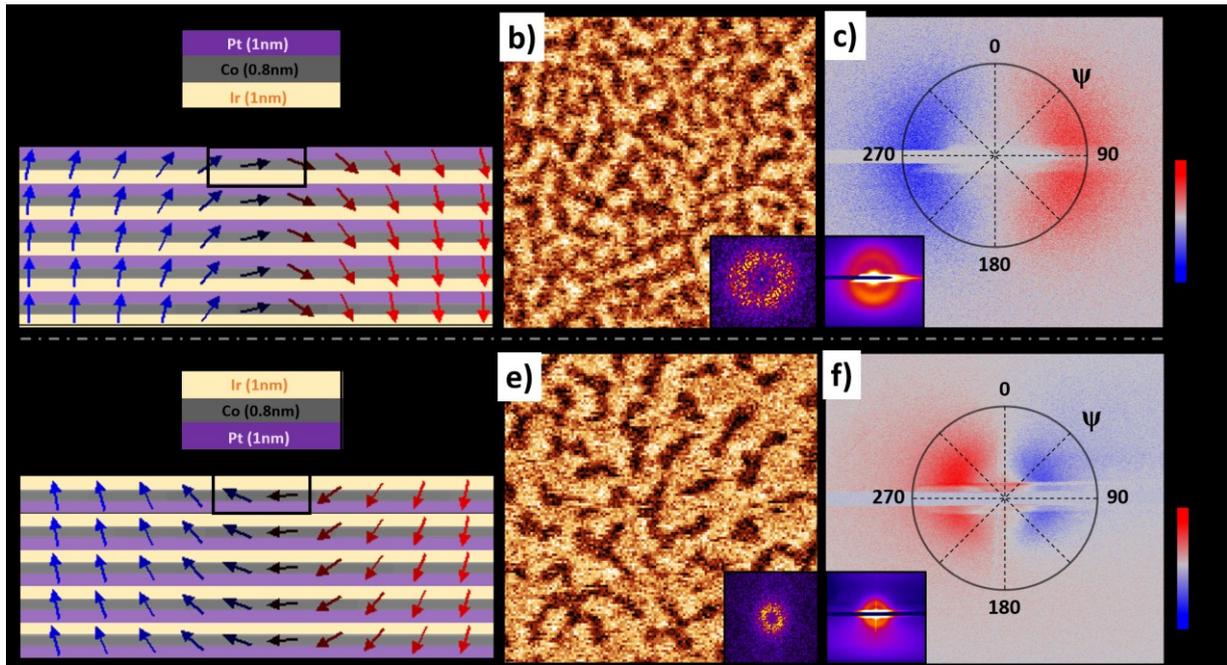

*Figure 3 : Comparison of two samples with inverted stacking //[Ir(1)/Co(0.8)/Pt(1)]x5 (a-c) and //[Pt(1)/Co(0.8)/Ir(1)]x5 (d-e). The corresponding magnetization textures are obtained by micromagnetic simulations based on material parameters (a & d). **(b & e)** MFM images with labyrinth-shape domains and their Fourier transform in inset. **(c & f)** Normalized circular dichroism signal ($I_{CL}$-$I_{CR}$)/($I_{CL}$+$I_{CR}$) measured on sample a (e) and sample b (f). The diffraction sum images are shown in inset, confirming again that the diffraction corresponds to the magnetic domains.*

In conclusion, we investigated the circular dichroism in the X-ray resonant magnetic scattering to determine chiral magnetic distribution, *i.e. its type (Neel or Bloch) and its magnetic chirality* in thin films presenting a strong DMI. In this work, we apply XRMS to the prototypical Pt/Co system, evidencing Néel type domain walls as well as their chirality depending on the stacking order. However, this method can be applied to any type of magnetic materials, either in static or in time resolved mode, and will be useful to investigate, for instance, metallic mutlilayers based on Co or Fe. Besides, further analysis associated to modelling might be performed to get some quantitative information such as DM magnitude and even anisotropies. The power of this method is its unambiguous conclusion regarding the DMI sign independently of other magnetic parameters. In a broader perspective, Circular Dichroism in x-ray scattering appears to be a unique tool for studying the type and the chirality of the magnetic phase of any magnetic material. Similar investigations could be profitably extended to different types of systems showing a non collinear magnetic ordering such as skyrmion lattices or conical/helical phases stabilized by the Dzyaloshinskii-Moriya interaction as well as more complex systems such as the recently observed antiskyrmion lattices [37].

Ackownledgments:

SoGraph (ANR-15-GRFL-0005) and MAGicSky (European Union Grant, No. FET-634 Open-665095) are acknowledged for financial support.

[1] S. Heinze, K. von Bergmann, M. Menzel, J. Brede, A. Kubetzka, R. Wiesendanger, G. Bihlmayer, and S. Blügel, Nat. Phys. **7**, 713 (2011).
[2] I. Dzyaloshinsky, J. Phys. Chem. Solids **4**, 241 (1958).
[3] T. Moriya, Phys. Rev. **120**, 91 (1960).
[4] X. Z. Yu, Y. Onose, N. Kanazawa, J. H. Park, J. H. Han, Y. Matsui, N. Nagaosa, and Y. Tokura, Nature **465**, 901 (2010).
[5] A. Fert, Mater. Sci. Forum **59**, 439 (1990).
[6] A. L. Balk, K.-W. Kim, D. T. Pierce, M. D. Stiles, J. Unguris, and S. M. Stavis, Phys. Rev. Lett. **119**, 077205 (2017).
[7] A. V. Khvalkovskiy, V. Cros, D. Apalkov, V. Nikitin, M. Krounbi, K. A. Zvezdin, A. Anane, J. Grollier, and A. Fert, Phys. Rev. B **87**, 020402 (2013).
[8] A. Thiaville, S. Rohart, É. Jué, V. Cros, and A. Fert, EPL Europhys. Lett. **100**, 57002 (2012).
[9] P.-H. Jang, K. Song, S.-J. Lee, S.-W. Lee, and K.-J. Lee, Appl. Phys. Lett. **107**, 202401 (2015).
[10] J. Sampaio, A. V. Khvalkovskiy, M. Kuteifan, M. Cubukcu, D. Apalkov, V. Lomakin, V. Cros, and N. Reyren, Appl. Phys. Lett. **108**, 112403 (2016).
[11] E. Jal, V. López-Flores, N. Pontius, T. Ferté, N. Bergeard, C. Boeglin, B. Vodungbo, J. Lüning, and N. Jaouen, Phys. Rev. B **95**, (2017).
[12] B. Pfau, S. Schaffert, L. Müller, C. Gutt, A. Al-Shemmary, F. Büttner, R. Delaunay, S. Düsterer, S. Flewett, R. Frömter, J. Geilhufe, E. Guehrs, C. M. Günther, R. Hawaldar, M. Hille, N. Jaouen, A. Kobs, K. Li, J. Mohanty, H. Redlin, W. F. Schlotter, D. Stickler, R. Treusch, B. Vodungbo, M. Kläui, H. P. Oepen, J. Lüning, G. Grübel, and S. Eisebitt, Nat. Commun. **3**, 1100 (2012).
[13] I. Mihai Miron, G. Gaudin, S. Auffret, B. Rodmacq, A. Schuhl, S. Pizzini, J. Vogel, and P. Gambardella, Nat. Mater. **9**, 230 (2010).
[14] L. Liu, C.-F. Pai, Y. Li, H. W. Tseng, D. C. Ralph, and R. A. Buhrman, Science **336**, 555 (2012).
[15] C. Moreau-Luchaire, C. Moutafis, N. Reyren, J. Sampaio, C. A. F. Vaz, N. Van Horne, K. Bouzehouane, K. Garcia, C. Deranlot, P. Warnicke, P. Wohlhüter, J.-M. George, M. Weigand, J. Raabe, V. Cros, and A. Fert, Nat. Nanotechnol. **11**, 444 (2016).
[16] H. Yang, A. Thiaville, S. Rohart, A. Fert, and M. Chshiev, Phys. Rev. Lett. **115**, 267210 (2015).
[17] M. Sacchi, N. Jaouen, H. Popescu, R. Gaudemer, J. M. Tonnerre, S. G. Chiuzbaian, C. F. Hague, A. Delmotte, J. M. Dubuisson, G. Cauchon, and others, in *J. Phys. Conf. Ser.* (IOP Publishing, 2013), p. 072018.
[18] N. Jaouen, J.-M. Tonnerre, G. Kapoujian, P. Taunier, J.-P. Roux, D. Raoux, and F. Sirotti, J. Synchrotron Radiat. **11**, 353 (2004).
[19] S. L. Zhang, G. van der Laan, and T. Hesjedal, Nat. Commun. **8**, 14619 (2017).
[20] S. L. Zhang, G. van der Laan, and T. Hesjedal, Phys. Rev. B **96**, (2017).
[21] H. A. Dürr, E. Dudzik, S. S. Dhesi, J. B. Goedkoop, G. Van der Laan, M. Belakhovsky, C. Mocuta, A. Marty, and Y. Samson, Science **284**, 2166 (1999).
[22] A. Fert, N. Reyren, and V. Cros, Nat. Rev. Mater. **2**, 17031 (2017).
[23] J. P. Hannon, G. T. Trammell, M. Blume, and D. Gibbs, Phys. Rev. Lett. **61**, 1245 (1988).
[24] J. P. Hill and D. F. McMorrow, Acta Crystallogr. A **52**, 236 (1996).
[25] G. van der Laan, Comptes Rendus Phys. **9**, 570 (2008).
[26] M. Blume and O. C. Kistner, Phys. Rev. **171**, 417 (1968).
[27] J. F. Pulecio, A. Hrabec, K. Zeissler, R. M. White, Y. Zhu, and C. H. Marrows, ArXiv161106869 Cond-Mat (2016).
[28] N. Romming, C. Hanneken, M. Menzel, J. E. Bickel, B. Wolter, K. von Bergmann, A. Kubetzka, and R. Wiesendanger, Science **341**, 636 (2013).
[29] J.-P. Tetienne, T. Hingant, L. J. Martínez, S. Rohart, A. Thiaville, L. H. Diez, K. Garcia, J.-P. Adam, J.-V. Kim, J.-F. Roch, I. M. Miron, G. Gaudin, L. Vila, B. Ocker, D. Ravelosona, and V. Jacques, Nat. Commun. **6**, 6733 (2015).

[30] Y. Dovzhenko, F. Casola, S. Schlotter, T. X. Zhou, F. Büttner, R. L. Walsworth, G. S. D. Beach, and A. Yacoby, ArXiv161100673 Cond-Mat (2017).
[31] G. Chen, S. P. Kang, C. Ophus, A. T. N'Diaye, H. Y. Kwon, R. T. Qiu, C. Won, K. Liu, Y. Wu, and A. K. Schmid, Nat. Commun. **8**, 15302 (2017).
[32] M. Belmeguenai, J.-P. Adam, Y. Roussigné, S. Eimer, T. Devolder, J.-V. Kim, S. M. Cherif, A. Stashkevich, and A. Thiaville, Phys. Rev. B **91**, 180405(R) (2015).
[33] H. S. Körner, J. Stigloher, H. G. Bauer, H. Hata, T. Taniguchi, T. Moriyama, T. Ono, and C. H. Back, Phys. Rev. B **92**, 220413(R) (2015).
[34] J. Torrejon, J. Kim, J. Sinha, S. Mitani, M. Hayashi, M. Yamanouchi, and H. Ohno, Nat. Commun. **5**, (2014).
[35] A. Authier, in *Int. Tables Crystallogr. Vol. B Reciprocal Space* (Springer, 2006), pp. 534–551.
[36] A. Vansteenkiste, J. Leliaert, M. Dvornik, M. Helsen, F. Garcia-Sanchez, and B. Van Waeyenberge, AIP Adv. **4**, 107133 (2014).
[37] A. K. Nayak, V. Kumar, T. Ma, P. Werner, E. Pippel, R. Sahoo, F. Damay, U. K. Rößler, C. Felser, and S. S. P. Parkin, Nature **548**, 561 (2017).